# SENSEMAKING IN CYBERSECURITY INCIDENT RESPONSE: THE INTERPLAY OF ORGANIZATIONS, TECHNOLOGY, AND INDIVIDUALS

*Research in Progress*

Lakshmi, Ritu, School of Computing and Information Systems, The University of Melbourne, Melbourne, Australia, rkudallur@student.unimelb.edu.au

Naseer, Humza, Centre for Future Enterprise, Queensland University of Technology, Brisbane, Australia, humza.naseer@qut.edu.au

Maynard, Sean, School of Computing and Information Systems, The University of Melbourne, Melbourne, Australia, sean.maynard@unimelb.edu.au

Ahmad, Atif, School of Computing and Information Systems, The University of Melbourne, Melbourne, Australia, atif@unimelb.edu.au

## Abstract

*Sensemaking is a critical activity in organizations. It is a process through which individuals ascribe meanings to events which forms the basis to facilitate collective action. However, the role of organizations, technology and individuals and their interaction in the process of sensemaking has not been sufficiently explored. This novel study seeks to address this gap by proposing a framework that explains how the interplay among organizations, technology and individuals enables sensemaking in the process of cybersecurity incident response. We propose that Organizations, Technology, and Individuals are the key components that interact in various ways to facilitate enactment, selection and retention activities (Sensemaking activities) in Incident Response. We argue that sensemaking in Incident Response is the outcome of this interaction. This interaction allows organizations to respond to cybersecurity incidents in a comprehensive manner.*

## 1 Introduction

Cybersecurity Incident Response (IR) is a crucial part of an organization's multi-layered approach towards protecting its business information assets (Ahmad et al., 2019). It emerged as a paradigm of cybersecurity management in response to the transient and unpredictable nature of threats faced by organizations (Baskerville et al., 2014). While identification of new threat types and responses has always been a constant challenge (Mitropoulos et al., 2006), the current pace of changes to the threat environment and increased occurrences of sophisticated, targeted attackers has intensified the pressure on incident response teams to protect organizational assets.

IR, is a complex, nuanced activity. This is because responders are required to interpret, analyse, and prioritize large volumes of information while collaborating with a diverse array of organizational stakeholders (H. Naseer et al., 2021; Werlinger et al., 2010). Swift and impactful decisions must be taken under these challenging circumstances to ensure that organizations sustain minimal damage from the cybersecurity incident. While organizations attempt to regulate the process by implementing planned activities and well-defined roles and responsibilities from best-practice guidelines (Cichonski, 2012; Tøndel et al., 2014), the unpredictable and emergent nature of incidents necessitates quick decision-making often in the face of voluminous and conflicting information and time constraints (A. Naseer et al., 2021).

It is here that we leverage sensemaking theory to understand how responders make sense in IR as sensemaking is well suited to study decision-making under challenging circumstances (Weick, 1993). Sensemaking is a critical activity in organizations and is a process through which individuals ascribe meanings to events which forms a basis to facilitate collective action (Brown et al., 2015).
Applying a sensemaking perspective to IR, we argue that sensemaking is the outcome of the interplay between Organizations, Technology, and Individuals (OTI). Sensemaking is made possible only due to the specific ways in which the OTI components interact along the response process. Consequently, we seek to explore the underlying mechanisms of the same by asking the following research question:

> *How does the interplay of Organization, Technology and individuals enable sensemaking in the process of cybersecurity incident response?*

To answer this question, we apply key components of Sensemaking Theory to Incident Response (IR) and develop a conceptual framework of sensemaking in IR. We begin with an overview of Sensemaking Theory and how it supports the development of our framework. This is followed by how Sensemaking has been addressed in IR. We then propose our framework in section 4, the components of which served as sensitizing concepts for the subsequent data collection. The paper concludes with the current status of the study and the potential implications of applying sensemaking lens to the process of Incident Response.

## 2      Theoretical Background

### 2.1   Making sense of sensemaking

Sensemaking is an interpretive, social constructivist approach to understanding how people cope with an equivocal world (Brown et al., 2015). It occurs when people attempt to understand events that are novel, ambiguous, or confusing by ascribing meanings to generate explanatory possibilities of what might be happening (Brown et al., 2015; Maitlis & Christianson, 2014). As a consequence of the process, intersubjective meaning is created to facilitate decision-making. While there are several definitions of sensemaking within the literature, we adopt the following definition which describes it in terms of several key properties – Sensemaking is social, retrospective, grounded in identity construction, enactive and involves construction of plausible narratives (Weick et al., 2005).

Sensemaking can occur in episodes triggered by an unplanned violation of expectations or in certain instances, be planned via "deliberate initiatives" where organizational actors are asked to make sense (Sandberg & Tsoukas, 2015). Once triggered, sensemaking occurs through a closed loop, overlapping set of enactment, selection and retention activities (Maitlis & Christianson, 2014). Enactment is the process whereby individuals identify and bracket portions of their experience for closer attention (Sandberg & Tsoukas, 2015). Once events are recognized, actors explicate the interpretive work performed and impose labels which "suggest(s) plausible acts of managing, coordinating and distributing" (Weick et al., 2005). Selection involves individuals working together to select a contextually rational explanation by "applying heuristic rules and repeatedly cycling through them" (Westwood & Clegg, 2003). It involves constructing a narrative that provides a rational explanation for seemingly conflicting evidence while being sufficient to initiate action. Retention involves the acquisition and storage of the contents of successful sensemaking and represents the changes that occur as a by-product of sensemaking activities (Brown et al., 2015). The "public" face of retention is reflected in changes to structures, routines and other material aspects of the enacted environment whereas the "private" face is reflected within the causal maps and beliefs that are internalized by the individuals involved.

Technology, however, has been largely absent from this discourse (Mesgari & Okoli, 2019; Sandberg & Tsoukas, 2015). While technology has been acknowledged as an enabler, we argue that it is a critical component of the sensemaking process due to the following reasons – first, technology itself is viewed as being equivocal i.e. a source of stochastic events that trigger sensemaking. Second, it influences the set of cognitive models that can be developed and the ways in which organizational actors understand and represent events associated with it (Weick, 1990). Third, it enables the discursive processes of sensemaking by supporting various communication practices. In addition to a technology void, sensemaking studies have also been criticized for a lack of understanding of the larger context within which it occurs (Sandberg & Tsoukas, 2020). While a large body of research linking institutional theory to sensemaking is available (Jensen et al., 2009; Weber & Glynn, 2006), the examination of the sensemaking process as the interaction among organizations, technology and individuals has not yet been conducted.

## 2.2 Cybersecurity Incident Response

Cybersecurity Incident Response (IR) is the process of responding to events that threaten the security of enterprise assets (Ahmad et al., 2019). It is a well-established process for which best practice guidelines are available from organizations such as NIST, ISO etc. (Cichonski, 2012). While organizations regulate the process by implementing these guidelines, the emphasis is on post-incident learning which helps fine-tune the process over time (Werlinger et al., 2007). The overall phases in the process include preparation, identification, containment, eradication, recovery, and follow-up (Ahmad et al., 2015). As part of this, responders identify incidents, perform diagnoses, collaborate with other stakeholders, and take actions for recovery (Werlinger et al., 2010). Responders rely heavily on using technology to perform IR activities. For example, tools such as system and network activity logging, security information and event management (SIEM) software and intrusion detection systems (IDS) are used to identify incidents, knowledge management systems are used for collaboration and analytics tools are used for incident diagnosis (H. Naseer et al., 2017, 2018; Tøndel et al., 2014). Technology is germane to Incident Response.

IR is standardized in terms of the expected activities that responders perform and hence most IR research is geared towards improving the efficiency and effectiveness of the response process (H. Naseer, Maynard, et al., 2016). Several studies approach IR as a series of action-events that occur when incidents are detected (E.g. Werlinger, Botta and Beznosov, 2007; Line, 2013; Tøndel, Line and Jaatun, 2014). This allows for the identification of bottlenecks along the process for which appropriate counter mechanisms can be implemented. For example, Werlinger, Botta and Beznosov (2007), while analysing how security practitioners respond to incidents, identify data collection as a critical task and describe impediments to the same. Studies have also attempted to understand the mental models developed by the responders and how it affects the subsequent diagnosis with the goal of facilitating better decision-making in IR. For example, Friedberg et al., (2015) describe methods for advanced data correlation to help detect Advanced Persistent Threats. Another stream of studies has approached IR as a problem of information flow, coordination, and control for effective response. For example, Sundaramurthy et al., (2016) have examined SOCs (Security Operation Centres) from an Activity Theory lens to identify the flow of information between participants for effective response. However, theoretical foundations explaining the process of sensemaking in cybersecurity incidents remain unexamined. Very little is known about how the interplay among organization, individuals and technology enables sensemaking in the process incident response. To address this gap, we employ sensemaking as the theoretical lens to explain how organizations, technology and individuals interact with each other to make sense of cyber incidents.

## 2.3 Sensemaking in Cybersecurity Incident Response

While sensemaking is endemic to any organizational activity, very few studies have applied the sensemaking lens to cybersecurity Incident response (E.g. Van der Kleij et al. 2017) . The dominant lens to study the IR process so far has been empiricist and transactional – attending more to the ergonomics, collaboration and problem structuration aspects and less on the decision maker's unfolding experience at the centre of the activity and the contextual factors that influence it. Despite the degree of planning involved, incidents are inherently unpredictable, necessitating swift decisions to be made under time constraints with incomplete and often conflicting information. Qualities such as flexibility, adaptability, improvisation and creativity have been highlighted as critical in dealing with the unpredictable and emergent nature of incidents (Bartnes et al., 2016; Brilingaitė et al., 2020). Responders, while working towards creating codified knowledge, always work at the edge of codified knowledge - where previous experience might not necessarily help cope with current events.

It is here that applying a sensemaking lens allows us to go beyond the processual aspects of IR to reframe it as a social process that acknowledges the uncertainty in the process, influence of context and active authoring by contributing participants (Maitlis & Christianson, 2014). It brings to the forefront nuances of negotiations, priorities and meaning that are mediated by social, cultural and technological activity settings in contemporary business environments and helps us deal with the complexity of the same (Namvar et al., 2018).

While sensemaking literature views sensemaking primarily as a socio-cognitive and a socio-organizational activity (Griffith, 1999; Sandberg & Tsoukas, 2015), the context of Incident Management necessitates the adoption of a socio-technical perspective. Given that Incident Management primarily plays out in the technological environment of the organization, it becomes material to the subsequent sensemaking that occurs with it. Technology here is not a passive participant that is stable and deterministic (Orlikowski & Scott, 2008; Weick, 1990) but rather behaves as an active agent – reciprocal in its relationship with the people who use it to make sense (Mesgari & Okoli, 2019; Orlikowski & Scott, 2008; Weick, 1990).

Extant studies on sensemaking and technology within Organization Studies focus more on how people make *sense of* new technologies as opposed to how users make *sense with* them (Mesgari & Okoli, 2019). A similar line of inquiry has been followed within the Information System discipline (E.g. Berente et al., 2011; Jensen et al., 2009). Consequently, there is little understanding of technology-mediated sensemaking. Some promising research in that direction involves investigating the role of technology artefacts within sensemaking (Yeow & Chua, 2020). However, given the embeddedness of technology and sensemaking in all organizational processes, it becomes essential to examine the interaction of these elements in an organizational setting. Hence we employ sensemaking as the theoretical lens to explain how organizations, technology and individuals interact with each other to make sense of cyber incidents.

## 3 Our Framework

Based on the analysis and integration of sensemaking and IR literatures, we develop a framework (see Figure 1) that explains how the interplay among organizations, technology and individuals enables sensemaking in the process of cybersecurity incident response. Sensemaking is the outcome of this interaction. Table 1 presents the definitions of key concepts in the framework.

### 3.1 Cyber Threat Environment

The cyber threat environment is the combination of external and internal environment of the organization from which cyber threats emerge due to the malicious activity of cyber threat actors (Desouza et al., 2020; Sundaramurthy et al., 2016). These threat actors are constantly inventing new tools and techniques and are getting better at identifying gaps and unknown vulnerabilities in the organization's security. The intention of cyber threat actors to conduct such activity is to compromise the security of the enterprise asset by altering its availability, integrity, or confidentiality (Ahmad et al., 2019; H. Naseer, Shanks, et al., 2016). In order to make sense of cyber threats and incidents that emerge from the cyber threat environment, organizations employ sensemaking process of enactment, selection and retention. Below we explain how organizations make sense of cybersecurity incidents through the interplay of OTI which, in turn, helps them respond to cybersecurity incidents in a comprehensive manner.

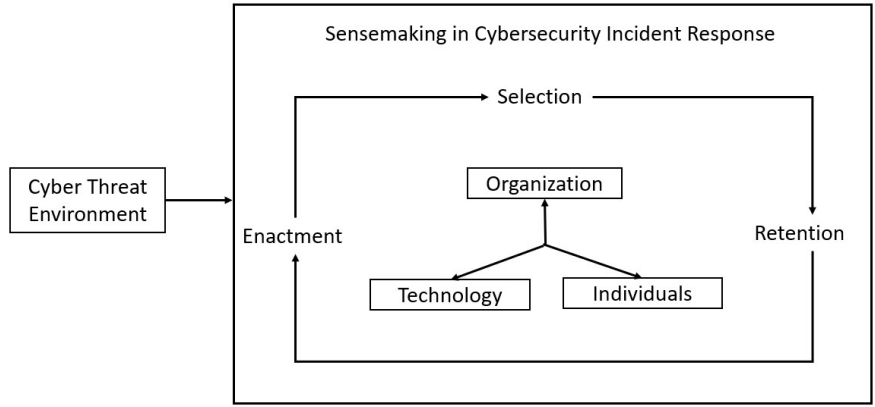

*Figure 1.     Framework of Cybersecurity Incident Response Sensemaking*

| Concept | | Definition | Reference |
|---|---|---|---|
| Cyber threat environment | | The external and internal environment of the organization from which Cybersecurity threats emerge | Sundaramurthy et al. 2016 |
| Sensemaking in cybersecurity incidents | Enactment | The act of individuals perceiving clues from their environment. | Weick, 1988; Weick et al., 2005 |
| | Selection | Process of individuals generating and negotiating intersubjective meaning to facilitate action. | Holt and Cornelissen 2014; Weick et al. 2005 |
| | Retention | Learning outcome of sensemaking that guides future rounds of enactment and selection | Jennings and Greenwood 2003; Weick 1979; Weick et al. 2005 |
| Organization | | The entity whose information assets needs to be protected. | Ahmad et al. 2015; Sundaramurthy et al. 2016 |
| Technology | | The set of tools, software and applications used to facilitate the Incident Response Process | Sundaramurthy et al. 2016; Werlinger et al. 2009 |
| Individuals | | The individuals participating in Incident Response Sensemaking – E.g. Responders, Subject Matter Experts | Ahmad et al. 2015; Sundaramurthy et al. 2016 |

*Table 1.     Definitions of concepts in the research framework*

## 3.2   Sensemaking in Incident Response

### 3.2.1   Enactment

Enactment is the process whereby individuals identify and bracket portions of their experience for closer attention on the basis of preconceptions (Weick et al., 2005). An example of this within Incident Response would be a responder noticing an alert on the intrusion detection system. Organizations enable enactment by imposing their beliefs over what constitutes an incident and how it is to be handled (Weber & Glynn, 2006). By setting up policies on incident classification and handling, driving the choices of tools and technologies implemented and setting up organizational structures to deal with incidents, organizations set up the IR processes. The individual responders operate within this framework which, over time, results in behavioural patterns that become sedimented as the institution (Weber & Glynn, 2006). Thus, the individual responder's behaviour is controlled by the institution.

Technology makes enactment possible by channelling the raw data from the environment (e.g. network traffic patterns) and making salient cues available for noticing by responders. Thus, a responder's ability to recognize and classify an event as an incident is mediated by technology in its ability to support collection and presentation of environmental cues (Tøndel et al., 2014; Werlinger et al., 2010). However, this influence of technology is possible only within the bounds of the organizational IR process. For example, an alert on the Intrusion Detection System by itself will not mean anything unless interpreted within the context of the larger IR process. An interaction of Organizations and Technology is a necessary condition for enactment to occur.

Cues noticed by the responder is mediated by technology and the organizational context. Factors such as the responders' experience, skill and role may enable them to do so (Werlinger et al., 2010). By designating the individual with the role of a responder, organizations drive expected behaviour regarding the perception of events as incidents (Jennings & Greenwood, 2003; Sandberg & Tsoukas, 2015). For example, an actor designated the role of a responder is more likely to notice and interpret the threat alerts appropriately as opposed to an actor whose role as an IT Manager may lead them to misjudge the severity (Weber & Glynn, 2006). The subsequent interpretation of what the alert means is influenced in a similar manner. Thus, all the OTI components need to interact for enactment to be made possible.

### 3.2.2 Selection

Selection involves individuals generating and negotiating intersubjective meaning via participation in socio-political processes (Holt & Cornelissen, 2014). There are two key changes that occur in the transition from enactment into selection. First, the articulation of the event by the individual actor which explicates individual perceptions thus offering actors categories and concepts to facilitate action (Weick, 2009). Second, there is a change from an individual-cognitive to a collective-discursive form of sensemaking. This implies that sensemaking is no longer within people's minds but is now required to be socially negotiated.

Within incident response, this would take the form of responders collaborating with experts to construct a narrative of how the incident occurred and its impact. This is made possible by the organizations which set up labels for classification (E.g. Minor vs Major), procedures to be invoked and reporting structures. Appropriate social feedback is set up in terms of roles and institutionalized expectations so that the selection processes are regulated (Weber & Glynn, 2006). However, organizations cannot operate these procedures without individuals' participation. Thus, individuals and organizations are required to interact via organizational decision structures that dictate who (e.g. First Responder vs Manager) and how (e.g. Meetings with Key Members of the team) decisions are to be taken.

Technology mediates the selection process by being able to integrate diverse information and presenting information for users to be able to make decisions. For example, tools such as knowledge management systems may be used by individuals to share information, IDM and Analytics help responders make better decisions and support the narratives constructed with evidence. The ability of a tool to be combined with other technologies to generate meaningful reports also helps facilitate selection activities (Tøndel et al., 2014; Werlinger et al., 2010). However, this is made possible only within the organizational context of IR where technology is used for goal-oriented tasks by individuals. Individuals influence the selection processes in terms of their roles (is the actor a designated Decision-Maker? – e.g. Manager), the amount of influence (does the actor have a large network?) and their expertise (do other individuals defer to the actor for information? e.g. First

Responder vs Manager). However, for individuals to influence selection processes, organizations need to endow them with the power to do so by way of assigning roles and creating processes to be followed.

### 3.2.3 Retention

Retention can be conceptualized as the learning that occurs as part of sensemaking. Retention in incident response takes the form of learning that is embodied in changes to the organizational processes and structures, individual causal maps, priorities, and attitudes (Colville et al., 2012). An example of retention in Incident Response is the changes in the policy and reporting structures that may occur as a result of post-incident learning. However, this occurs only when organizations set up appropriate retention activities (Hove et al., 2014; Tøndel et al., 2014). Technology supports the retention process by way of its ability to retain changes (e.g. changes to configuration, detection rules), the richness of information supported and its ability to recall in a timely manner (time lag vs instantaneously) (Paul, 2006). However, these retained changes do not have meaning unless organizations guide it towards goal-oriented learning for the purposes of responding to incidents.

Individual retention can be visualized as updates to the causal maps and priorities of responders (Sandberg & Tsoukas, 2015). For example - A post incident review process highlights some communication gaps between the IT and Business which results in changes to the reporting structure. The new reporting structure implements the creation of a report that details the Business impact of the incident that is to be sent to the CFO. This results in changes to the causal maps within responders who are more likely to visualize the business impact while making sense of subsequent attacks.

Thus, the specific ways in which the OTI components interact results in sensemaking within Incident response. While the components have individual contributions, we argue that sensemaking is dependent upon the three interacting together.

## 4 Proposed Method

To study the interplay of OTI components, we have taken on a two-phase data collection approach combining expert interviews and multiple case studies. At the time of writing, we have completed phase one which is composed of expert interviews with Cybersecurity professionals (in particular, Security Operation Centre team, cybersecurity managers, and cybersecurity analysts). Participants selected were from both the consulting and corporate domains.

Our aim is to collect data from as many participants as we can until data saturation is achieved. Specifically, we sought participants who possess a wide repertoire of in-depth knowledge and experiences in incident response. We are asking them questions on how Organizations, Technology and Individuals (OTI) interact with each other during the Incident Response process to make sense of cybersecurity incidents. The objective of the expert interviews is to refine our framework and ensure that a comprehensive set of sensitising concepts informed by both practice and the prior literature are identified. In addition, inferences drawn from the expert interviews will be used in the next phase.

In phase two of the research, we will conduct multiple in-depth case studies with organizations that have a mature incident response function. We have already identified and have established contact with three organizations from banking and finance sector. Our aim is to conduct 5-7 case studies in total. The procedures for case studies will follow Yin's (2018) methods. Each case will be conducted and analysed individually, and cross-case conclusions will be drawn to modify and refine the framework developed in the first phase. Multiple case studies with specific organizational context will add depth to the research by providing rich insight on (1) mechanisms through which OTI components interact with each other, and (2) the dynamic capabilities that emerge from making sense of cybersecurity incidents through OTI interplay over time. We intend to follow Gioia, Corley and Hamilton's (2013)'s methodology, a grounded theory technique, to refine and further develop concepts in our framework that mesh theory and data.

## 5  Conclusion, Contributions and Limitations

This paper seeks to understand how sensemaking occurs in the cybersecurity critical process of Incident Response. By identifying the critical components of IR – Organizations, Technology and Individuals (OTI) we argue that sensemaking is the outcome of the interplay of these components. While individuals perform sensemaking activities, the interplay of organizations and technology is what makes sensemaking possible.

This study provides potential contributions to the literature in several ways. First, it advances current knowledge by developing a conceptual understanding of how incident responders make *sense* in the IR process. Second, this study addresses the technology gap by including it as one of the key components implicated in sensemaking process. Third, it develops a theoretical framework that explains how enactment, selection and retention (key sensemaking activities) occurs in IR via the interaction of OTI components. Fourth, it contributes to the cybersecurity IR literature by providing detailed definitions of concepts and their relationships in the framework, grounded in the sensemaking theory.

This study also offers several useful implications for practitioners. First, it highlights the importance of employing sensemaking processes of enactment, selection and retention in IR practice to make *sense* of cyber threats and incidents that emerge from the cyber threat environment. Second, the proposed research framework considers interaction of OTI components as the key enabler of sensemaking activities in the IR process. This perspective is important as it presents IR process as the interplay among socio-technical systems and how they coalesce to detect and respond to cybersecurity incidents in a comprehensive manner. Finally, the continuous practice of sensemaking activities in IR process will help cybersecurity managers to make informed and swift decisions related to cybersecurity incidents.

As qualitative work, the proposed empirical study will ultimately be limited by the quality of the expert interviews and multiple case studies data we obtain. While we have established excellent access to informants from a broad range of organizations for our expert interviews and in-depth case studies, the generalization of our findings would be difficult as they are context specific. Nonetheless, we make a bold step in trying to understand how the interplay among organization, technology and individuals enables sensemaking in IR, a question that is of importance to organizations of all sizes and levels of establishment. As the application of sensemaking lens in cybersecurity incident response is still limited, we expect future research to not only refine and evolve our concepts but also further elaborate sensemaking as a theory and explore its application in cybersecurity.